\documentclass[prl,twocolumn,showpacs]{revtex4}
\usepackage{amssymb}
\usepackage{graphicx,graphics,color,times,slashed}

\begin{document}

\title{An Experimental Proposal to Test Dynamic Quantum Non-locality \\
with Single-Atom Interferometry}
\author{Shi-Liang Zhu$^1$, Zheng-Yuan Xue$^1$, Dan-Wei Zhang$^1$, and
Lu-Ming Duan$^2$}
\affiliation{$^1$Laboratory of Quantum Information Technology, ICMP and SPTE, South China
Normal University, Guangzhou, China\\
$^2$Department of Physics and MCTP, University of Michigan, Ann Arbor,
Michigan 48109, USA}

\begin{abstract}
Quantum non-locality based on the well-known Bell inequality is of
kinematic nature. A different type of quantum non-locality, the
non-locality of the quantum equation of motion, is recently put
forward with connection to the Aharonov-Bohm effect [Nature Phys.
6, 151 (2010)]. Evolution of the displacement operator provides an
example to manifest such dynamic quantum non-locality. We propose
an experiment using single-atom interferometry to test such
dynamic quantum non-locality. We show how to measure evolution of
the displacement operator with clod atoms in a spin-dependent
optical lattice potential and discuss signature to identify
dynamic quantum non-locality under a realistic experimental
setting.
\end{abstract}

 \maketitle

Non-locality dramatically exemplified in the Einstein-Podolsky-Rosen (EPR)
paradox \cite{EPR} is a fundamental concept of quantum mechanics that
distinguishes it from classical physics. Quantum non-locality based on the
EPR correlation was later formulated into an experimentally testable result
known as the Bell inequality \cite{Bell}. Quantum non-locality based on the
EPR correlation and the Bell inequality has been verified in many
experiments involving different physical systems \cite{Genovese}.

Recently, a different type of quantum non-locality, the non-locality of the
quantum equation of motion \cite{Aharonov83}, implied in the famous
Aharonov-Bohm effect \cite{Aharonov59}, was put forward by Popescu \cite%
{Popescu}. Non-locality based on violation of the Bell inequality comes from
the Hilbert space structure of quantum mechanics and thus is purely
kinematic, while non-locality implied in the Aharonov-Bohm effect is from
non-locality of quantum equations of motion in the Heisenberg picture and
thus is of dynamic nature \cite{Popescu}. Another significant difference
between these two kinds of non-localities is that non-locality from the EPR
correlation is assumed to be an exclusive quantum property of two or more
well separated but entangled particles, while the dynamic quantum
non-locality (DQNL) considered by Aharonov et al. and Popescu can be
demonstrated even with evolution of a single particle in a superposition
state of two distinct locations. The evolution of the displacement operator
provides an explicit example to clearly show the DQNL \cite{Popescu},
however, experiment is still lacking in this direction due to the difficulty
to measure the displacement operator.

In this paper, we propose a feasible experiment using cold atoms in a
spin-dependent optical lattice potential to test the DQNL. We figure out a
configuration where the DQNL\ inherent in the Heisenberg equation leads to a
detectable signal qualitatively different from that of the corresponding
local (classical) evolution equation, and propose a method to directly
measure evolution of the displacement operator in the real experimental
system. The required ingredients in this proposed experiment, such as the
double well optical lattice and the spin-dependent movement of a particle,
have all been realized in previous experiments \cite{Mandel,Bloch,Lee}, and
thus the proposal well fits with the status of the current technology.

Before explaining the proposal, first we briefly recall the concept of DQNL
elaborated in Ref. \cite{Popescu}. The Schrodinger equation describing
evolution of the wave function of a quantum system is always a local
differential equation, however, the wave function by itself is not directly
observable. To see the DQNL, one needs to look at the Heisenberg equations
which describe evolution of observable physical quantities. The Heisenberg
equation for the displacement operator $\hat{D}^{Q}$ provides an example to
explicitly show this kind of DQNL for a single particle \cite{Popescu}. The
displacement operator $\hat{D}^{Q}$ is defined as $\hat{D}^{Q}\equiv \exp [i%
\hat{p}L/\hbar ]$ with $\hat{p}$ being the momentum operator of the
particle. This operator shifts the particle by a finite distance $L$. For
simplicity, we consider a one-dimensional situation where the Hamiltonian of
the particle is given by $H=\frac{\hat{p}^{2}}{2m}+V(x)$ with $m$ being the
mass of the particle and $V(x)$ being the potential. For this system, the
classical and quantum equations of motion of the displacement operator are
quite different \cite{Popescu}. In classical mechanics, we can apply the
chain rule for differentiation of a function, and evolution of the quantity $%
D^{C}\equiv \exp [ipL/\hbar ]$ is given by

\begin{equation}
\frac{d{D}^{C}}{dt}=\frac{de^{ipL/\hbar }}{dp}\frac{dp}{dt}=\frac{L}{i\hbar }%
e^{ipL/\hbar }\frac{dV(x)}{dx},  \label{D_S}
\end{equation}%
which is a local differential equation. However, quantum mechanically, the
displacement operator $\hat{D}^{Q}$ is governed by the Heisenberg equation,
which leads to

\begin{equation}
\frac{d\hat{D}^{Q}}{dt}=\frac{1}{i\hbar }\left[ \hat{D}^{Q},H\right] =\frac{1%
}{i\hbar }[V(x+L)-V(x)]\hat{D}^{Q},  \label{D_Q}
\end{equation}%
where we have used $e^{i\hat{p}L/\hbar }V(x)=V(x+L)e^{i\hat{p}L/\hbar }$.
This evolution equation is clearly nonlocal as the time derivative of the
quantity depends on the potential at two distinct (and possibly remote)
locations $x$ and $x+L$.

To demonstrate this kind of DQNL, we need to figure out a configuration
where the classical and the quantum evolution equations (1) and (2) for the
displacement operator show clear qualitative difference. We also need to
find a method to measure the displacement operator in real experimental
systems. The evolution operator $\hat{D}^{Q}$ is non-Hermitian, so it is not
directly observable. However, we can look at the real and imaginary parts of
$\hat{D}^{Q}$, and they correspond to observable quantities and still
satisfy nonlocal evolution equations in quantum mechanics. To have a
configuration that manifests the DQNL represented by Eq. (2), we consider a
particle confined in one dimension with a double-well potential, as shown in
Fig. 1(a-c). The two potential wells are identical, except that the bottom
of one of the wells may be shifted with that of the other well by a constant
energy $\Delta $. For classical particles in either of these wells, they see
identical force and cannot tell the difference of the wells. The local
dynamic equation should be independent of the energy shift $\Delta $.
However, for quantum particles in a superposition state, the evolution of
the displacement operator can sense this nonlocal constant energy shift $%
\Delta $. To be explicit, let us assume that the potential $V(x)$ around the
two minima $\pm L/2$ can be described by the identical harmonic trap, with $%
V_{1}(x)=\left( m\omega ^{2}/2\right) \left( x+L/2\right) ^{2},\ \
V_{2}(x)=\left( m\omega ^{2}/2\right) \left( x-L/2\right) ^{2}+\Delta ,$
where $\omega $ is the characteristic trap frequency. Let $|\Phi (x)\rangle $
denote an eigenstate of the harmonic trap (for convenience, it can be taken
as the ground state). We take the initial state of the particle at time $t=0$
as the following superposition of two localized wave packets,
\begin{equation}
|\Psi (x,0)\rangle =\left[ |\Phi (x+L/2)\rangle +|\Phi (x-L/2)\rangle
e^{i\theta }\right] /\sqrt{2},  \label{Initial_state}
\end{equation}%
where $\theta $ is an arbitrary initial phase difference. The size of the
wave packet $|\Phi (x)\rangle $ at each well, estimated by $\sqrt{\hbar
/m\omega }$, is assumed to be significantly smaller than $L$ so that the
overlap $\int \Phi ^{\ast }(x+L/2)\Phi (x-L/2)dx\approx 0$. For this state,
the quantum Heisenberg equation (2) for the displacement operator directly
gives

\begin{equation}
\left\langle \frac{d\hat{D}^{Q}}{dt}\right\rangle =-i\omega _{d}\left\langle
\hat{D}^{Q}\right\rangle ,  \label{Dot_D}
\end{equation}%
where $\omega _{d}=\Delta /\hbar .$ It has the straightforward solution
\begin{equation}
\langle \hat{D}^{Q}\left( t\right) \rangle =\left\langle \hat{D}%
^{Q}(0)\right\rangle e^{-i\omega _{D}t}=\left( 1/2\right) e^{i\theta
-i\omega _{d}t}.  \label{D_Q1}
\end{equation}%
So the evolution of the quantum displacement operator is sensitive to the
nonlocal constant energy shift $\Delta $. In contrast, for classical
particles with the dynamic equation (1), even if they are distributed over
the two wells, as long as the distribution function in each well is
symmetric with respect to the trap bottom (which is the case for the state
shown in Eq. (3)), the average force $\left\langle \frac{dV(x)}{dx}%
\right\rangle $ is always zero and $\left\langle \frac{d{D}^{C}}{dt}%
\right\rangle =0$. So, as expected, the classical dynamic equation cannot
sense the nonlocal constant energy shift and there is a qualitative
difference in the measurement outcomes for the classical and the quantum
evolution equations for the displacement operator.

The phenomenon discussed above is closely related to the scalar
Aharonov-Bohm effect \cite{sAB,sAB2}: in classical physics, the constant
energy shift does not lead to any physical difference as long as the force
is identical in space. However, for a quantum particle in a superposition
state, it can sense the constant energy shift at two remote locations even
if the force is strictly zero at any point of the particle's trajectory.
This is similar again to the conventional Aharonov-Bohm\ effect \cite%
{Aharonov59}, where a quantum charged particle senses a nonzero constant
vector potential while the electromagnetic force is zero at any point of the
particle's trajectory.

To prepare the particle in a superposition state of a double well potential
and to directly measure the displacement operator, we propose to use an
optical lattice potential to control cold atoms to fulfill all the
requirements. We consider dilute atomic gas in an optical lattice with the
average filling number per lattice site much less than $1$, so the atomic
interaction is negligible and we just have many independent copies of
single-particle dynamics. To generate a superposition state over different
lattice sites, one may use a double well lattice along the $x$ direction,
with the potential $V\left( x\right) =-V_{1}\sin ^{2}\left( kx/2+\varphi
\right) -V_{2}\sin ^{2}\left( kx\right) $ ($V_{1},V_{2}>0$) from two
standing wave laser beams with the wave vector $k=2\pi /\lambda \equiv \pi
/L $ \cite{Bloch,Lee}. Initially, we set the phase $\varphi =0$ and $%
V_{2}/V_{1}=0$ so that we have only a single lattice $V_{1}$\ and the atom
is in the ground state of this lattice well. By adiabatically tuning up the
ratio $V_{2}/V_{1}$ to the region with $V_{2}\gg V_{1}$, each lattice site
splits into two as shown in Fig. 1 (a-b), and the atomic state adiabatically
follows the ground state configuration and evolves into an equal
superposition state of the two wells in the form of Eq. (3) with $\theta =0$%
. After we have prepared this initial state, we quickly (within a time scale
$t_{\delta }$) tune $\varphi $ and $V_{1}$ so that $\varphi =\pi /4$ and $%
V_{1}=\Delta /2\ll V_{2}$, and look at evolution of the displacement
operator under this double well potential with a constant energy shift $%
\Delta $ (Fig. 1c). Under the lattice $V_{2}$, the potential well is approximated by a
harmonic trap with the trapping frequency $\omega =2\sqrt{V_{2}E_{r}}/\hslash $,
where $E_{r}=\hbar ^{2}k^{2}/2m$ is the atomic recoil energy. We require the
time scale $t_{\delta }$ to satisfy the condition $\omega ^{-1}\ll t_{\delta
}\ll \hslash /\Delta $ so that on the one hand, we do not generate motional
excitations in each well, and on the other hand, $t_{\delta }$ is negligible
compared with the evolution time scale (of the order of $\hslash /\Delta $) of the displacement operator.

To demonstrate the nonlocal dynamics of the displacement operator $\hat{D}%
^{Q}$ shown in Eq. (5) under this double well lattice, we need to measure $%
\hat{D}^{Q}$ after a controllable time delay $t$. The displacement
operator does not correspond to a simple physical quantity, and it
is not easy to measure it directly in experiments. To overcome
this problem, we make use of the internal (spin) states of the
atoms. We show in the following that a Ramsey type of experiment
in the internal state, together with a spin-dependent movement of
the lattice potential, gives a direct measurement of the
expectation value of the displacement operator. The atoms have
different hyperfine states, and we use two of them, denoted by
effective spin $|\uparrow \rangle $ and $|\downarrow \rangle $,
respectively. The atoms are initially assumed to be in the state
$|\uparrow \rangle $. To measure the displacement operator after
an evolution time $t$ ($t\sim\hslash /\Delta $), we take the
following four steps as illustrated in Fig. 1(d-f): (i) first, we
apply a $\pi /2$-pulse within a time much shorter than $\hslash
/\Delta $ to the atomic internal state so that the atomic state
transfers to $\left[ \left( |\uparrow \rangle +|\downarrow \rangle
\right) /\sqrt{2}\right] |\psi \left( t\right) \rangle $, where
$|\psi \left( t\right) \rangle$ denotes the atomic motional state
at time $t$. (ii) Second, we apply a spin dependent shift to the
lattice potential with the corresponding unitary operation
described by $U=|\uparrow \rangle \left\langle \uparrow
\right\vert \hat{D}_{L/2}^{Q}+|\downarrow \rangle \left\langle
\uparrow \right\vert \hat{D}_{-L/2}^{Q}$. This kind of operation
has been realized
before in experiments to demonstrate controlled atomic collisions \cite%
{Mandel}. This operation needs to be done in a time scale $t_{\delta }$
which satisfies $\omega ^{-1}\ll t_{\delta }\ll \hslash /\Delta $ so that
the shift of the lattice does not generate motional excitations in each
well. After this step, the atomic state becomes $\left( |\uparrow \rangle
\hat{D}_{L/2}^{Q}+|\downarrow \rangle \hat{D}_{-L/2}^{Q}\right) |\psi \left(
t\right) \rangle /\sqrt{2}$. (iii) After the spin-dependent lattice shift,
we apply another $\pi /2$-pulse (within a time negligible compared with $%
\hslash /\Delta $) to the atomic internal state which transfers the atomic
state to $|\psi _{f}\rangle =\left[ \left( |\uparrow \rangle +|\downarrow
\rangle \right) \hat{D}_{L/2}^{Q}+\left( |\downarrow \rangle -|\uparrow
\rangle \right) \hat{D}_{-L/2}^{Q}\right] |\psi \left( t\right) \rangle /2$.
(iv) Finally, we measure the total atom number $N_{\downarrow }$ in the spin
down state minus the total number $N_{\uparrow }$ in the spin up state. This
number difference is proportional to the probability difference $%
P_{\downarrow }-P_{\uparrow }$ for the state$|\psi _{f}\rangle $, which is
given by%
\begin{equation}
P_{\downarrow }-P_{\uparrow }=Re\left[ \langle \Psi \left( t\right) |\hat{D}%
_{L}^{Q}|\Psi \left( t\right) \rangle \right] .
\end{equation}%
The imaginary part of the expectation value $\langle \Psi \left( t\right) |%
\hat{D}_{L}^{Q}|\Psi \left( t\right) \rangle $ can be measured in a similar
way. The only difference is that in the step (i) we add a relative phase $i$
to the $\pi /2$-pulse which transfers spin $|\uparrow \rangle $ to the state
$\left( |\uparrow \rangle +i|\downarrow \rangle \right) /\sqrt{2}$. The
final state is then modified to $|\psi _{f}\rangle =\left[ \left( |\uparrow
\rangle +|\downarrow \rangle \right) \hat{D}_{L/2}^{Q}+i\left( |\downarrow
\rangle -|\uparrow \rangle \right) \hat{D}_{-L/2}^{Q}\right] |\psi \left(
t\right) \rangle /2$ with the probability difference $P_{\downarrow
}-P_{\uparrow }=Im\left[ \langle \Psi \left( t\right) |\hat{D}_{L}^{Q}|\Psi
\left( t\right) \rangle \right] $.

\begin{figure}[tbp]
\label{Fig1} \includegraphics[width=7.5cm,height=5cm]{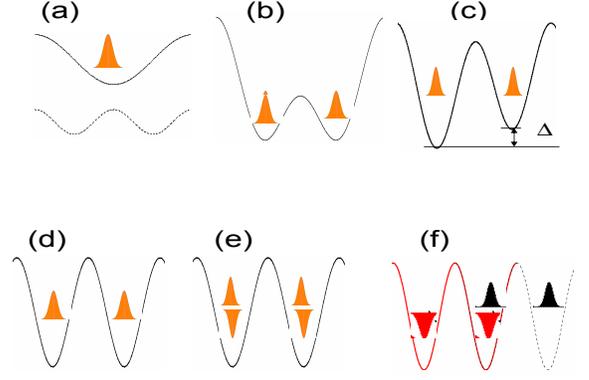}
\caption{(Color online) Illustration of the experimental steps for
initial state preparation and for detection of the displacement
operator. Figs. (a)-(c) show the steps to prepare a non-local
superposition state (given by Eq. (3)) in a double-well lattice.
After adiabatic preparation of this superposition state, a bias
potential $\Delta$ is tuned on within a time scale
$t_\protect\delta$ (specified in the text) to start evolution of
the displacement operator. Figs. (d-f) show the steps to measure
the displacement operator after a certain evolution time $t$.
Right before the measurement, the bias potential $\Delta$ is
turned off, so we have a regular optical lattice (d). The atom is
then transfer to an equal superposition of two spin components
with a $\protect\pi/2$-pulse (e). After a spin-dependent shift of
the optical lattice (f), followed by another
$\protect\pi/2$-pulse, we measure the population difference in
these two spin components, and this difference gives directly the
expectation value of the displacement operator. }
\end{figure}

In the above, we have shown how to measure the expectation value of the
displacement operator for cold atoms in an optical lattice. The DQNL
indicates that the real (imaginary) part of this expectation value
oscillates with the evolution time $t$ as $\cos (\omega _{d}t)$ (-$\sin
(\omega _{d}t)$) as predicted by Eq. (5), which is sensitive to the nonlocal
constant energy shift $\Delta =\hslash \omega _{d}$. This signal
distinguishes it from the corresponding classical case where $\left\langle {D%
}^{C}\right\rangle $ is independent of $\Delta $ and shows no oscillation
with time $t$. For real experiments in an optical lattice, however, there is
inevitably a global harmonic trap potential (taking the form of $%
V_{t}=m\omega _{t}^{2}x^{2}/2$ in the $x$ direction) which could complicate
the situation \cite{Bloch,Lee}. The measured atom number difference $%
N_{\downarrow }-N_{\uparrow }$ involves average of the probability
difference $P_{\downarrow }-P_{\uparrow }$ over all the independent
double-well potentials. Due to the global trap $V_{t}$, the probability
difference $P_{\downarrow} -P_{\uparrow }$ in each double well potential
oscillates with slightly different frequencies, and one needs to check
whether an average over the whole lattice will wash out the oscillation
signal. For this purpose, we simulate in Fig. (2) the averaged signal for a
typical experimental configuration. The averaged probability difference is
given by
\begin{equation}
\left\langle P_{\downarrow }-P_{\uparrow }\right\rangle =\frac{1}{2N_{l}}%
\sum_{j}\cos (\Delta _{j}t/\hbar )e^{-\gamma t},
\end{equation}%
where $\Delta _{j}=\Delta +\delta _{j}$ with $\delta _{j}\approx m\omega_t
^{2}Lx_{j}$ ($L=\lambda /2,$ and $x_{j}$ denotes the coordinate of the
center of the double wells). The summation of $j$ is over all the occupied
double wells (with number $N_{l}$) in the global harmonic trap. To better
model the experimental situation, we also add a phenomenological decay $%
e^{-\gamma t}$ to each oscillation term which corresponds to a nonzero
dephasing rate $\gamma $ inevitable in reality. Under typical experimental parameters we
have damped oscillations as shown in Fig. 2. The signal is still clearly observable
in this case. In the frequency domain, the spectrum centers at the energy
shift $\Delta /\hslash $ that is independent of the experimental
imperfection discussed above.

\begin{figure}[tbp]
\label{Fig2} \includegraphics[width=7.5cm,height=6cm]{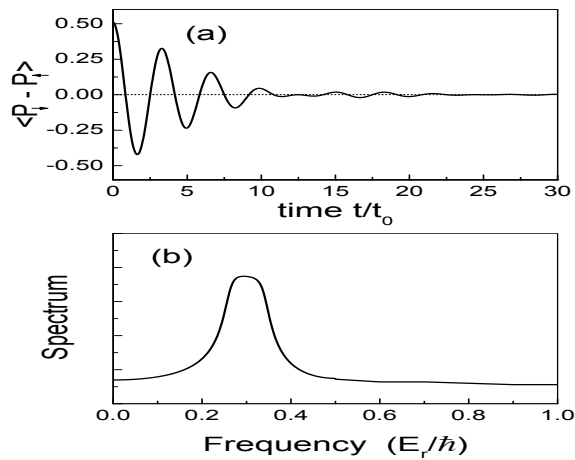}
\caption{Simulated experiential signal in an inhomogeneous optical
lattice with a global harmonic trap. (a): Averaged population
difference of the two spin components as a function of evolution
time (in the unit of $t_{0}=\hbar /E_{r}$) of the displacement
operator. The average is taken over $40$
occupied double wells in an global harmonic trap with the trap frequency $%
\protect\omega_t =2\protect\pi \times 50$ Hz. Other parameters include $L=%
\protect\lambda/2=400$nm, the atomic mass $m=1.45\times 10^{-25}$ kg for $%
^{87}$Rb atoms, and the bias potential $\Delta =0.3E_{r}$. A dephasing rate $%
\protect\gamma =0.1E_{r}/\hbar$ is assumed (see Eq. (7). (b) Fourier
transform of the signal in Fig. (a). Instead of a sharp line at the bias
potential ($\Delta/\hbar$,) the curve has a broad peak due to the broadening
from average in the inhomogeneous global trap and the nonzero dephasing
rate. However, the peak is still centered at the bias potential $\Delta/\hbar
$. }
\end{figure}

Before ending the paper, we briefly discuss the requirements for the
relevant experimental parameters. To assure locality, we assume the wave
packet overlap between different wells is negligible. This overlap is
estimated by $e^{-L/l_{0}}$, where $l_{0}=\sqrt{\hbar /m\omega }$ is the
size of the wave packet in each well and $L=\lambda /2$ is the distance
between the wells. For $Rb^{87}$ atoms in an optical lattice with $\lambda
=800$ $nm$, $\omega =2\sqrt{V_{2}E_{r}}\sim 2\pi \times 42$ $kHz$ and $%
l_{0}\sim 52$ $nm$ for \ a typical lattice barrier $V_{2}=35$ $E_{r}$,
the condition $e^{-L/l_{0}}\sim e^{-7.6}\ll 1$ is well satisfied. During the
state preparation and the detection of the displacement operator, we require
the operation time $t_{\delta }$ to satisfy $\omega ^{-1}\ll t_{\delta }\ll
\hslash /\Delta $. If we take $\Delta \sim 0.3E_{r}\sim 0.025\omega $ and $%
t_{\delta }\sim 8\omega ^{-1}\sim 30$ $\mu s$, the motional excitations
estimated by the Landau-Zener formula is small and all the requirements seem
to be reasonable with the current experimental technology.

In summary, we have proposed a feasible experiment using cold atoms in an
optical lattice to test the DQNL associated with evolution of the
displacement operator. The DQNL is different from and complementary to the
kinetic quantum non-locality represented by the Bell inequalities.
Similar to tests of the Bell inequalities, an experimental test of the DQNL could
shed new light on our understanding of fundamentals of quantum mechanics.

The work was supported by the NSF of China (No 10974059), the State Key
Program for Basic Research of China (Nos.2006CB921801 and 2007CB925204), the
MURI, and the DARPA program.

\end{document}